\documentclass[notitlepage,12pt]{gvbarticle}
\usepackage{amsmath}
\usepackage{times}
\usepackage{mhequ}			 
\usepackage{epsfig}
\usepackage{sub_JP}
\newtheorem{theorem}{Theorem}[section]

\newtheorem{corollary}[theorem]{Corollary}
\newtheorem{definition}[theorem]{Definition}
\newtheorem{lemma}[theorem]{Lemma}
\newtheorem{proposition}[theorem]{Proposition}
\newtheorem{remark}[theorem]{Remark}

\newenvironment{proof}[1][Proof]{\textit{#1.} }{\ \rule{0.5em}{0.5em}\par}

\newcommand{\supt}{{\displaystyle\sup_{t\geq0}}}

\renewcommand{\epsilon}{\varepsilon}

\def\L{{\rm L}}

\newcommand{\ed}{e}
\def\newpsi{\psi}
\def\newchi{\chi}
\def\psitzero{\psi_{a}}
\def\chitzero{\chi_{a}}
\def\tzero{1}
\def\hatpsitzero{\tilde\psi_a}
\def\deltat{\delta}
\def\KSL{${\rm KS}_L$}
\def\eg{{\it e.g.},}
\def\ie{{\it i.e.},}
\def\Y{Y}
\def\ca{c}
\def\d{{\rm d}}
\def\pint{{\rm P}\kern-0.9em\int}

\def\HALF{{\textstyle\frac{1}{2}}}

\def\ximinus{\varphi}

\def\mindelta{\min(\delta_*\kern-0.05em,\,t)}
\def\Ltwo{2}
\def\Lone{1}
\begin{document}

\title{Non-Vanishing Profiles for the Kuramoto-Sivashinsky Equation on
the Infinite Line}
\author{G.~van Baalen\inst{1}, J.-P.~Eckmann\inst{1}\fnmsep\inst{2}}
\institute{D\'epartement de Physique Th\'eorique, Universit\'e de
Gen\`eve \and Section de Math\'ematiques, Universit\'e de Gen\`eve
}
\titleindent=0.65cm
\maketitle
\thispagestyle{empty}
\pagestyle{MHheadings}
\begin{abstract}We study the Kuramoto-Sivashinsky equation on the 
infinite line with initial conditions having arbitrarily large limits
$\pm \Y$ at 
$x=\pm\infty$. We show that the solutions have the same limits for all
positive times. This implies that an
attractor for this equation cannot be defined in $\L^\infty$. 
To prove this, we consider
profiles with limits at $x=\pm\infty$, and show that initial conditions
$\L^2$-close to such profiles lead to solutions which remain $\L^2$-close to
the profile for all times. Furthermore, the difference between these solutions and the
initial profile tends to $0$ as $x\to\pm\infty$, for any fixed time
$t>0$.
Analogous
results hold for $\L^2$-neighborhoods of periodic stationary solutions.
This implies that profiles and periodic stationary solutions partition the
phase space into mutually unattainable regions.
\end{abstract}
\section{Introduction}
The Kuramoto-Sivashinsky equation
\begin{equs}
\dot{\mu}(x,t)=-\mu''''(x,t)-\mu''(x,t)-\HALF (\mu^2)'(x,t)~,~~~~~
\mu(x,0)=\mu_0(x)
~,
\label{eqn:KS}
\end{equs}
is an interesting model for stabilization mechanisms of very indirect
type. It can be considered on a finite interval of length $L$ with
periodic boundary conditions (\KSL) or on the infinite line ($\rm
KS_{\rm {\infty }}$). In both cases, one can formally multiply the
equation with $\mu$ and integrate, leading to (after integration by
parts)
\begin{equs}
\HALF \partial _t \int \d x\,\mu^2=
\int \d x\, \left (\mu'\right)^2 (x,t)-\int \d x\,\left (\mu''\right)^2 (x,t)-\HALF 
\int \d x\,\mu\cdot (\mu^2)'.
\end{equs}
Note that the last term vanishes identically since it equals $\int
(\mu^3)'/6$ and thus, surprisingly, the non-linearity does not
contribute directly to the decay of initial conditions with large
$\L^2$ norm, in contrast to equations like the Ginzburg-Landau
equation \cite{Mielke2002} which derive from a potential.\footnote{On
the other hand, as we shall see, it is precisely this feature which
allows for $\L^2$ bounds which grow only exponentially in time.}
Clearly, on the other hand, without the non-linear term, the equation
is unstable. It has been shown
\cite{NST1985,Ilyashenko1992,CEES1993a,Goodman1994} that for finite
$L$, the equation \KSL~has an attractor in $\L^2$ as well as in
$\L^{\infty}$, whose radius is finite (known to be bounded by
$L^{8/5}$, resp. $L^{48/25}$, see \eg~\cite{CEES1993a,CEES1993b}).
Numerical experiments seem to indicate that these bounds should in
fact be extensive (with at worst ${\cal O}(\log L)$ corrections), \ie~
$L^{1/2}$ for the $\L^2$ radius and $L^{0}$ for $\L^{\infty}$ . Here,
we will show that this conjecture is wrong for initial conditions in
$\L^\infty({\bf R}) $, since we shall construct initial data $\mu_\Y $
satisfying $\lim_{x\to\pm \infty }\mu_\Y(x)=\pm \Y $ and {\em for
which the corresponding solution\footnote{the well known Bunsen flame
fronts (see \cite{Michelson1996}).} $\mu$ satisfies
\begin{equs}\lim_{x\to\pm \infty}\mu(x,t)=\pm\Y 
\label{eqn:limit}
\end{equs} 
for all $t>0$.} The behavior of (\ref{eqn:limit}) is similar to what
happens for the diffusion equation, where initial conditions with
different limits at $\pm\infty $ also maintain this property as time
increases, see \eg~\cite{ColletEckmann1992}. Note that due to the
Galilean invariance of (\ref{eqn:KS}), the extensivity conjecture
${\cal O}(L^0)$ for the attractor in $\L^{\infty}$ is trivially wrong
without a requirement ruling out constants -- which are global
solutions of (\ref{eqn:KS}) -- as admissible initial data. The usual
restriction to break the Galilean symmetry is to require the initial
datum to be an odd function (see e.g.~\cite{CEES1993a}). For our
purpose however, it is enough to assume that the limits at
$x=\pm\infty$ are of opposite sign and equal magnitude.

What do we learn from (\ref{eqn:limit}) ? Basically, it shows, that if
there is ever to be a definition of attractor for ${\rm KS_\infty}$ it
must contain constraints on the initial condition which are much
stronger than just being in $\L^\infty $ (plus asymmetry and arbitrary
regularity). Rather, if there is any hope to define a bounded
attractor, it would have to come from a condition which says that the
initial condition looks everywhere ``like'' those well-known
\cite{Manneville1990} patterns one encounters in numerical
simulations. In the absence of a technique replacing localization as
in \cite{ColletEckmann1990, MielkeSchneider1995} this seems currently
impossible to achieve. Note however that it {\em is} known
\cite{Michelson1986} that periodic stationary solutions are
universally bounded. The present paper obtains more information on the
structure of the phase space, if not on an eventual attractor, by
showing that periodic stationary solutions and profiles divide
naturally the phase space into mutually unattainable regions.

Our proof of (\ref{eqn:limit}) is based on the following simple idea.
First of all, constants are clearly stationary solution of ${\rm
KS_\infty}$. Furthermore, (\ref{eqn:KS}) has a one parameter family of
explicit (albeit unbounded) solutions of the form
\begin{equs}
\mu(x,t)=\frac{bx}{1+bt}~,
\end{equs}
with $b>0$, showing that positive constant slopes are rotated
clockwise. Our starting point consists in combining these two special
solutions by taking as an initial condition the function
\begin{equs}
\psi_a(x)=a \arctan \left ( {x}\right )~,
\end{equs}
where $a= 2\Y /\pi$. The `middle' of this function is like the
constant slope example (with $b=a$) while for large $x$ it reaches
very quickly $\pm\Y $. It is therefore natural to assume that
\begin{equs}
\psi_a(x,t)=a \arctan \left (\frac {x}{1+at}\right )
\end{equs}
is a good approximate profile for $t>0$. In fact, while the definition
of $\psi_a(x,t)$ is suggestive, as long as we only have bounds, and
not small bounds in $\L^2$, we may, and will, work with the {\em
fixed} function $\psi_a=\psi_a(x)$ that is, with a profile which is
not changing in time. Our main result is that {\em if the initial
condition $\mu_0$ is $\L^2$-close to $\psi_a(x)$ and the difference
decays at infinity then the solution remains $\L^2$-close to
$\psi_a(x)$, and the difference still decays at infinity for all
$t>0$.} This result also holds if $\psi_a$ is replaced by
$\psi_{per}$, with $\psi_{per}$ a periodic and analytic stationary
solution of (\ref{eqn:KS}) as constructed in \cite{Michelson1986}.
Therefore, there exist solutions of ${\rm KS_\infty}$ which stay near
$\pm\Y $ at infinity for all times, and thus we have found a family of
large initial conditions whose evolution does not get smaller in
$\L^\infty $ as time goes to $\infty $.\footnote{Note that we do not
claim (and it quite probably is not true) that $\mu(x,t)-\psi_a(x,t)$
stays bounded in $\L^2$. We will rather see that it grows (quickly) in
$\L^2$. But the only thing which matters is that it remains in $\L^2$
and decays at infinity.} On the other hand, initial conditions which
behave asymptotically like periodic stationary solutions apart from
$\L^2$ corrections remain so for all times. Since the difference of
two periodic functions with different periods is {\em not} in $\L^2$,
this shows that the phase space naturally splits into disconnected
components. This last result is an extension of \cite{Grujic2000}.

The discussion above  suggests to consider the equation for
$\nu(x,t)=\mu(x,t)-\newpsi(x)$, which reads
\begin{equs}
\dot{\nu}=
-\nu''''-\nu''
-\HALF (\nu^2)'-(\nu\newpsi)'+\newchi~,
~~~~~\nu(x,0)=\nu_0(x)
\label{eqn:KSmodif}
\end{equs}
where 
\begin{equs}
\newchi\equiv-\newpsi''''-\newpsi''-\newpsi\newpsi'
~,
\end{equs} 
and $\lim_{x\to\pm\infty}\nu_0(x)=0$. We will consider
(\ref{eqn:KSmodif}) either with $\newpsi=\psitzero$ and corresponding
$\chitzero$, or with $\newpsi=\psi_{per}$ a periodic analytic
stationary solution for which $\newchi=\chi_{per}=0$. Instead of
$\psitzero$, we could have used the stationary profiles
(\ie~stationary solutions of (\ref{eqn:KS})) constructed in
\cite{Michelson1986}), or even the explicit one
\begin{equs}
\mu(x,t)=\frac{15}{361} \sqrt{209}
\left(-9 
\tanh\Big({\textstyle\frac{\sqrt{209}}{38}} x\Big)
+11\tanh\Big({\textstyle\frac{\sqrt{209}}{38}}x\Big)^3
\right)
\end{equs}
found by Kuramoto \cite{Kuramoto1976}. Note that these profiles are
uniformly bounded. The advantage of these choices would have been that
$\chitzero=0$, the disadvantage is the lack of explicit formulas, in
particular for the Fourier transform of the profiles. While adding an
inhomogeneous term to the equation, the choice of $\psitzero$ retains
the main properties of these stationary profiles, {\it e.g.} in terms
of analyticity. As is easily seen, high frequency modes are strongly
damped by (\ref{eqn:KS}) at the linear level. It is known (see
\eg~\cite{CEES1993b}) that solutions corresponding to periodic
antisymmetric initial condition in $\L^2([-L/2,L/2])$ become analytic
in finite time in a strip of finite width around the real axis. The
error term $\chitzero$ of the equation (see (\ref{eqn:KSmodif})) and
$\psitzero $ are analytic in the strip $|{\rm Im}~z|<1$ and uniformly
bounded in any smaller strip---these two facts are better seen in
Fourier space, since the Fourier transform $\hatpsitzero $ of
$\psitzero $ exists as a distribution and is given by
\def\alpha{a}
\begin{equs}
\hatpsitzero (k)=a\frac{\ed^{-|k|}}{k}~.
\label{eq:ft}
\end{equs}
\begin{definition}
Throughout, we denote by $A$ the operator $A=\sqrt{-\partial_x^2}$.
\end{definition}
\begin{remark}
We fix $\Y>0$, and we tacitly admit that all constants occurring
in the sequel may depend on $\Y$.
\end{remark}
\begin{theorem}\label{thm:main}
There are constants $\ca$ and $\beta>0$ such that the following holds.
For any initial condition $\nu(\cdot,0)$ with
$\nu_0\equiv\|\nu(\cdot,0)\|_{\Ltwo}<\infty $, the solution of
(\ref{eqn:KSmodif}) exists for all $t>0$ and
\begin{equs}
\supt~\ed^{-\beta t}
\|\nu(\cdot,t)\|_{\Ltwo}
\leq
\nu_0+\ca~.
\end{equs}
Furthermore, the flow is regularizing in the sense that there exist
constants $\delta_* >0$, $\gamma\geq\beta$, and $C<\infty $ such that
\begin{equs}
\sup_{t\geq0}
~
\ed^{-\gamma t}
\|\ed^{\mindelta A}\nu(\cdot,t)\|_{\Ltwo}\le\nu_0+C~.
\end{equs}
\end{theorem}
\begin{corollary}\label{cor:tendtozero}
For every $m=0,1,\dots$ there exists a constant $C_m$ such that 
\begin{equs}
|\partial_x^m\nu(x,t)|
\leq \frac{C_m}{\mindelta^m}\ed^{\gamma  t}~~\mbox{and}~~
\lim_{x\to\pm\infty}\partial_x^m\nu(x,t)=0
\end{equs}
for all $t>0$.
\end{corollary}
\begin{proof}
By Theorem \ref{thm:main} there exists a $C'$ such that
$\|\ed^{\mindelta A}\nu(\cdot,t)\|_{\Ltwo}\le C'\ed^{\gamma t}$. By
the Schwarz inequality,
\begin{equs}
\|A^m\tilde{\nu}(\cdot,t)\|_{\Lone}\leq
\left(\int{\rm
d}k~|k|^{2m}\ed^{-2\mindelta |k|}\right)^{1/2}
\|\ed^{\mindelta A}\nu(\cdot,t)\|_{\Ltwo}~.
\end{equs}
This immediately implies the first assertion
since $\sup_x
|\partial_x^m\nu(x,t)|\leq\|A^m\tilde{\nu}(\cdot,t)\|_{\Lone}$.
This bound also implies 
(by the Riemann-Lebesgue theorem) that
$\displaystyle\lim_{x\to\pm\infty}\partial_x^m\nu(x,t)=0$.
\end{proof}

\section{Functional Spaces, Estimates}
In this section, we collect a few straightforward bounds on the
function $\psitzero$ and the operator $A=\sqrt{-\partial_x^2}$. We
denote by ${\cal W}_{\tau,\deltat }$ the (Banach) space obtained by
completing ${\cal C}_0([\tau,\tau+\deltat ],{\cal C}_{0}^{\infty}({\bf
R}))$ in the norm $\sup_{t\in[\tau,\tau+\delta
]}\|\ed^{(t-\tau)A}\cdot\|_{\Ltwo}$, where, throughout, $\|\cdot\|_p$
is the $\L^p$ norm. We also denote by ${\cal B}_d\subset{\cal
W}_{\tau,\deltat }$ the open ball of radius $d$, centered on $0$ in
${\cal W}_{\tau,\deltat }$. The bounds of this section serve to
control the non-linear and mixed terms in Eq.(\ref{eqn:KSmodif}).
\begin{lemma}\label{lem:withpsi}
There is a $\delta _*>0$ such that
for all $t\in[\tau,\tau +\delta _*]$ one has
\begin{equs}
\|\ed^{(t-\tau)A}~\psitzero (\cdot)f(\cdot,t)\|_{\Ltwo}
\leq 2\pi~a~
\|\sqrt{1+A^2}~\ed^{(t-\tau)A}f(\cdot,t)\|_{\Ltwo}~.
\end{equs}
\end{lemma}
\begin{remark}
One can choose $\delta _*=1$ as will be seen from the proof. (This
value is related to the domain of analyticity of
$\psi_a$.)
\end{remark}
\begin{proof}
Define $F(x,t)=\ed^{(t-\tau)A}~\psitzero
(x)f(x,t)$. We write the Fourier transform of $F$ as
\begin{equs}
\tilde F(k,t)=\ed^{ (t-\tau )|k|}\int \d\ell\,
\tilde\psitzero(k-\ell) \tilde f(\ell,t)
~.
\end{equs}
Using (\ref{eq:ft}), we find (using principal values)
\begin{equs}
\tilde F(k,t)=a\,\ed^{(t-\tau )|k|}\int \d\ell\,
\frac{\ed^{- |k-\ell|}}{k-\ell} \tilde f(\ell,t)
~.
\end{equs}
Denote $g(x,t)=\ed^{(t-\tau)A}f(x,t)$, so that $\tilde
f(\ell,t)=\ed^{- (t-\tau )|\ell|}\tilde g(\ell,t)$. Rearranging
exponentials, we get
\begin{equs}
\tilde F(k,t)&=a\,\int \d\ell\,
\frac{\ed^{- |k-\ell|}}{k-\ell} \ed^{ (t-\tau
)|k|}\ed^{- (t-\tau )|\ell|}\tilde g(\ell,t) \cr
&=a\,\int \d\ell\,
\frac{\ed^{-|k-\ell|(1-(t-\tau))}}{k-\ell} \ed^{ (t-\tau
)\bigl(|k|-|\ell| -| k-\ell|\bigr)}\tilde g(\ell,t)~.\cr
\end{equs}
We decompose this as
\begin{equs}
\tilde{F}(k,t)&=
a\int
{\rm d}\ell ~
\frac{\ed^{-|k-\ell|(1-(t-\tau))}}{k-\ell }
~\tilde{g}(\ell ,t)+
a\int
{\rm d}\ell ~
~G(k,\ell ,t-\tau,(1-(t-\tau)))
~\tilde{g}(\ell ,t)~,
\end{equs}
where 
\begin{equs}
G(k,\ell ,\xi,\eta)=
\frac{\ed^{-\eta|k-\ell |}}{k-\ell }
~\left(\ed^{\xi(|k|-|\ell|-|k-\ell |)}-1\right)~.
\end{equs}
One checks easily, using the triangle inequality, that for $\xi\ge0$,
\begin{equs}
|G(k,\ell ,\xi,\eta)|\leq
|G(k,\ell ,\xi,0)|\leq\frac{\xi}{\sqrt{1+(\xi(k
-\ell ))^2}}~.
\end{equs}
Using  $\int
{{\rm d}k}({1+k^2})^{-1}=\pi$, we get
\begin{equs}
\|F(\cdot,t)\|_{\Ltwo}&\leq
\|\psitzero (\cdot)\ed^{(t-\tau)A}~f(\cdot,t)\|_{\Ltwo}+
a\sup_{\ell \in{\bf R}}\|G(\cdot,\ell ,t-\tau,\tzero +\tau)\|_{\Ltwo}
~\|\tilde{g}(\cdot,t)\|_{\Lone}\\
&\leq
\frac{a\pi}{2}\|\ed^{(t-\tau)A}~f(\cdot,t)\|_{\Ltwo}+
a\sup_{\ell\in{\bf R}}\|G(\cdot,\ell ,t-\tau,0)\|_{\Ltwo}
~\Big\|
\frac{\sqrt{1+A^2}}{\sqrt{1+A^2}}~\tilde{g}(\cdot,t)
\Big\|_{\Lone}\\
&\leq
\pi~a~\Big(\HALF +\sqrt{t-\tau}\Big)~
\|\sqrt{1+A^2}~\ed^{(t-\tau)A}~f(\cdot,t)\|_{\Ltwo}~\\
&\leq
2\pi~a
\|\sqrt{1+A^2}~\ed^{(t-\tau)A}~f(\cdot,t)\|_{\Ltwo}~,
\end{equs}
provided $\delta _*\le1$.
The proof of Lemma \ref{lem:withpsi} is complete.
\end{proof}

\begin{lemma}\label{lem:withpsiper}
Let $\psi_{per}$ be periodic of period $L$, let $q=\frac{2\pi}{L}$ and
assume that there exist constants $c_{per},\delta$ such that
$\sum_{m\in{\bf Z}}|\psi_{per,m}|\ed^{\delta q|m|}<c_{per}$, where
$\psi_{per,m}$ denotes the $m$-th Fourier coefficient of $\psi_{per}$,
then
\begin{equs}
\|\ed^{(t-\tau)A}\psi_{per}(\cdot)f(\cdot,t)\|_{\Ltwo}\leq
c_{per}~\|\ed^{(t-\tau)A}f(\cdot,t)\|_{\Ltwo}
\end{equs}
for all $t\in[\tau,\tau+\delta]$.
\end{lemma}
\begin{proof}
As in Lemma \ref{lem:withpsi}, we define
$F(x,t)=\ed^{(t-\tau)A}~\psi_{per}
(x)f(x,t)$. The Fourier transform of $F$ satisfies
\begin{equs}
\tilde F(k,t)=\ed^{ (t-\tau )|k|}\sum_{m\in{\bf Z}}
\psi_{per,m} f(k-qm)
~,
\end{equs}
so that
\begin{equs}
\|F(\cdot,t)\|_{\Ltwo}&\leq
\sum_{m\in{\bf Z}}
|\psi_{per,m}|\left(
\int{\rm d}k~\ed^{2(t-\tau)|k+qm|}
~|f(k)|^2\right)^{1/2}\\
&\leq
\|\ed^{(t-\tau)A}f(\cdot,t)\|_{\Ltwo}~
\sum_{m\in{\bf Z}}
|\psi_{per,m}|~\ed^{(t-\tau)q|m|}~,
\end{equs}
which completes the proof of the lemma since $0\leq t-\tau\leq\delta$.
\end{proof}

\begin{lemma}\label{lem:trilinear}
Let $\gamma>0$ and $\|(1+A^2)\ed^{\gamma
A}~f\|_{\Ltwo}+\|\ed^{\gamma A}~g\|_{\Ltwo}+\|\ed^{\gamma
A}~h\|_{\Ltwo}<\infty$. Then 
\begin{equs}
\left|
\int{\rm d}x \,\ed^{\gamma A}~f\cdot
\ed^{\gamma A}(gh)'\right|
\,\leq\,
\sqrt{\pi}
\|(1+A^2)\ed^{\gamma A}~f\|_{\Ltwo}
~\|\ed^{\gamma A}~g\|_{\Ltwo}
~\|\ed^{\gamma A}~h\|_{\Ltwo}~.
\end{equs}
\end{lemma}
\begin{proof}
Set $F=\ed^{\gamma A}~f$, $G=\ed^{\gamma A}~g$ and $H=\ed^{\gamma
A}~h$. Since $|k|-|k-\ell |-|\ell |\leq0$ by the triangle inequality,
we have
\begin{equs}
\left|
\int{\rm d}x\, \ed^{\gamma A}~f\cdot
\ed^{\gamma A}(gh)'\right|
&=
\left|
\int{\rm d}k
~ik\tilde{F}(k)
\int{\rm d}\ell 
~\ed^{\gamma(|k|-|k-\ell |-|\ell |)}
~\tilde{G}(k-\ell )
~\tilde{H}(\ell )
\right|~\\
&\leq
\Big\|\frac{\sqrt{1+A^2}}{\sqrt{1+A^2}}A\tilde{F}\Big\|_{\Lone}\,
\|G\|_{\Ltwo}\,\|H\|_{\Ltwo}\\
&\leq\sqrt{\pi}\|(1+A^2)F\|_{\Ltwo}\,
\|G\|_{\Ltwo}\,\|H\|_{\Ltwo}~,
\end{equs}
where we used again $\int{\rm d}k\,({1+k^2})^{-1}=\pi$.
\end{proof}
The following proposition estimates how close $\psi_a$ is to a
solution of ${\rm KS}_\infty $.
\begin{proposition}\label{lem:forchi}
Define
$\psitzero (x)=a\arctan(x)$ and let $\chitzero =-\psitzero''''
-\psitzero ''-\psitzero^{\vphantom'} \psitzero'$. 
Then, for $0\le\delta \le\delta _*\equiv\HALF$, one has
\begin{equs}
\sup_{t\in[0,\deltat ]}
\|\ed^{ tA}\chitzero \|_{\Ltwo}
\leq B~,
\end{equs}
for some $B$ depending only on $a$.
\end{proposition}
\begin{proof}
The Fourier transform of $\ed^{tA}\psi_a''''$ is of the form
$i^3a\ed^{t|k|} k^3\ed^{-|k|}$ so that for $t\le\delta $ we get
\begin{equs}
\|\ed^{tA}\psi_a''''\|_{\Ltwo}^2\le a^2 \int \d k\,
\ed^{2(\delta-1) |k|} k^6\le {\cal O}((1-\delta )^{-7})~.
\end{equs}
A similar bound holds for $\psi_a''$. The term $\ed^{tA} \psi_a
\psi_a'$ is bounded using Lemma \ref{lem:withpsi} with $\delta
_*=\HALF$ instead of 1 and $f=\psi_a'$. This yields a bound on the
square of the $\L^2$ norm which is of the form $a^2{\cal O}(1) \int \d
k\, (1+k^2) e^{2\delta |k|} e^{-2|k|}$, and combining the bounds
completes the proof.
\end{proof}

\section{The Local Cauchy Problem in $\L^2$}
In this section, we consider the local (in time) Cauchy problem 
\begin{equs}
\dot{\nu}=
-\nu''''-\nu''
-\HALF (\nu^2)'-(\nu\psitzero )'+\chitzero ~,
~~~~~\nu(x,\tau)=\nu_0(x)~,
\label{eqn:KSmodifrap}
\end{equs}
for (\ref{eqn:KSmodif}) with $\nu_0\in\L^2$. We will show, using a
contraction argument,  that it is
well posed on any time interval $t\in[\tau,\tau+\deltat ]$ with
$\deltat \le \min(\delta _*,C_*\|\nu_0\|_{\Ltwo}^{-2})$.\footnote{Note
that by our choice of $\delta _*$, we have $t-\tau \le \HALF$.} To
this end, we construct the map $\rho\mapsto{\cal F}(\rho)$ defined by
${\cal F}(\rho)=\xi$, where $\xi$ is the solution of
\begin{equs}
\dot{\xi}=
-\xi''''-\xi''
-\HALF (\xi\rho)'-(\xi\psitzero )'+\chitzero ~,
~~~~~\xi(x,\tau)=\nu_0(x)~,
\label{eqn:defF}
\end{equs}
and show that if $\deltat $ is sufficiently small ($\deltat
\sim\|\nu_0\|_{\Ltwo}^{-2}$) then ${\cal F}$ is a contraction in a
ball of radius $>\|\nu_0\|_{\Ltwo}$ in ${\cal W}_{\tau,\deltat }$.
Namely, let $f=\ed^{(t-\tau)A}\xi$ and $g=\ed^{(t-\tau)A}\rho$.
Multiplying (\ref{eqn:KSmodifrap}) with $f\ed^{(t-\tau)A}$,
integrating over the space variable and using the results of the
preceding section, we have
\begin{equs}
\HALF \partial_t\|f\|_{\Ltwo}^2&\leq
 \|A^{1/2}f\|_{\Ltwo}^2
-\|A^{2}f\|_{\Ltwo}^2
+\|Af\|_{\Ltwo}^2
+{\textstyle\frac{\sqrt{\pi}}{2}}
\|(1+A^{2})f\|_{\Ltwo}
\|f\|_{\Ltwo}\|g\|_{\Ltwo}\\
&\phantom{\leq}
+2\pi a
\|Af\|_{\Ltwo}
\|\sqrt{1+A^{2}}f\|_{\Ltwo}
+B\|f\|_{\Ltwo}~.
\label{eqn:thekey}
\end{equs}
The first term on the r.h.s.~comes from the time derivative of the
exponential $\ed^{(t-\tau )A}$, the second and third from the space
derivatives of $\xi$. The next term uses Lemma \ref{lem:trilinear},
while the last two use Lemma \ref{lem:withpsi} and Proposition
\ref{lem:forchi}, respectively. Then, we use the inequalities
\begin{equs}
\|Af\|_{\Ltwo}\le \|\sqrt{1+A^{2}}f\|_{\Ltwo}~,~~~~
\|\sqrt{1+A^{2}}f\|_{\Ltwo}^2\leq
\|(1+A^{2})f\|_{\Ltwo}
\|f\|_{\Ltwo}
\end{equs}
and get
\begin{equs}
\HALF \partial_t\|f\|_{\Ltwo}^2&\leq
 \|A^{1/2}f\|_{\Ltwo}^2
-\|A^{2}f\|_{\Ltwo}^2
+\|Af\|_{\Ltwo}^2+B\|f\|_{\Ltwo}
\\
&\phantom{\leq}
+\left({\textstyle\frac{\sqrt{\pi}}{2}}\|g\|_{\Ltwo}
+{2\pi a}\right)
\|(1+A^{2})f\|_{\Ltwo}
\|f\|_{\Ltwo}~.
\end{equs}
We also have
\begin{equs}
\|A^{1/2}f\|_{\Ltwo}^2
&\leq
\|f\|_{\Ltwo}^{3/2}\|A^2f\|_{\Ltwo}^{1/2}
\leq\frac{3}{4\epsilon_1^{1/3}}\|f\|_{\Ltwo}^2
+\frac{\epsilon_1}{4}\|A^2f\|_{\Ltwo}^2~,\\
\|Af\|_{\Ltwo}^2&\leq
\|f\|_{\Ltwo}\|A^2f\|_{\Ltwo}\leq
\frac{1}{2\epsilon_2}\|f\|_{\Ltwo}^2
+\frac{\epsilon_2}{2}\|A^2f\|_{\Ltwo}^2~,\\
X\|(1+A^2)f\|_{\Ltwo}&\leq
\frac{1}{2\epsilon_3}X^2
+\frac{\epsilon_3}{2}
\Big(\|A^2f\|_{\Ltwo}^2+2\|Af\|_{\Ltwo}^2+\|f\|_{\Ltwo}^2\Big)~,
\end{equs}
for all $X,\epsilon_i>0,i=1,2,3$. Using these inequalities with
sufficiently small $\epsilon_i$ shows that there is a positive
constant $c_*$ such that
\begin{equs}
\partial_t\|f\|_{\Ltwo}^2 \leq
(c_*+\|g\|_{\Ltwo}^2)
\|f\|_{\Ltwo}^2+
B^2~,
\end{equs}
from which we get that for all $\rho$ in the ball of radius $d$ in
${\cal W}_{\tau,\deltat }$, ${\cal F}(\rho)$ satisfies
\begin{equs}
\sup_{t\in[\tau,\tau+\deltat ]}
\|\ed^{(t-\tau)A}{\cal F}(\rho)\|_{\Ltwo}
\leq
\ed^{{\delta} (c_*+d^2)/2 }
\sqrt{\|\nu_0\|_{\Ltwo}^2+B^2}~.
\end{equs}
For all $d>\sqrt{\|\nu_0\|_{\Ltwo}^2+2B^2}$, there exists a $\deltat
={\cal O}(\|\nu_0\|_{\Ltwo}^{-2})$ such that ${\cal F}$ maps ${\cal
B}_d\subset{\cal W}_{\tau,\deltat }$ strictly into itself. On the
other hand $\ximinus={\cal F}(\rho_1)-{\cal F}(\rho_2)$ satisfies
\begin{equs}
\dot\ximinus =
-\ximinus ''''-\ximinus ''
-\HALF (\ximinus \bar{\rho})'
-\HALF (\bar{\xi}(\rho_{1}-\rho _{2}))'-(\ximinus \psitzero )'
~,
~~~~~\ximinus (x,\tau)=0~,
\end{equs}
where $\bar{\rho}=({\rho_1+\rho_2})/{2}$ and $\bar{\xi}=\bigl({\cal
F}(\rho_1)+{\cal F}(\rho_2)\bigr)/{2}$. Since $\bar{\rho}$ and
$\bar{\xi}$ are in ${\cal B}_d\subset{\cal W}_{\tau,\deltat }$,
similar arguments also show that for the same $d$ and $\deltat $ as
above
\begin{equs}
\sup_{t\in[\tau,\tau+\deltat ]}
\|\ed^{(t-\tau)A}({\cal F}(\rho_1)-{\cal F}(\rho_2))\|_{\Ltwo}
<
\sup_{t\in[\tau,\tau+\deltat ]}
\|\ed^{(t-\tau)A}(\rho_1-\rho_2)\|_{\Ltwo}
~,
\end{equs}
so that ${\cal F}$ is a contraction in ${\cal B}_d\subset{\cal
W}_{\tau,\deltat }$. Thus the sequence of approximating solutions
$\nu_{n+1}={\cal F}(\nu_n)$ converges to a unique solution of
(\ref{eqn:KSmodifrap}) in ${\cal B}_d\subset{\cal W}_{\tau,\delta}$.

Note that the results of this section also hold for the equation
\begin{equs}
\dot{\nu}=
-\nu''''-\nu''
-\HALF (\nu^2)'-(\nu\psi_{per})'~,
~~~~~\nu(x,\tau)=\nu_0(x)~.
\label{eqn:KSmodifper}
\end{equs}
It follows easily from \cite{Michelson1986} (see also \cite{Grujic2000})
that periodic stationary solutions $\psi_{per}$ of (\ref{eqn:KS}) satisfy
the hypotheses of Lemma \ref{lem:withpsiper}. The procedure is then exactly
the same, \ie~to show that the analog of the map ${\cal F}$ is a
contraction in ${\cal B}_d\subset{\cal W}_{\tau,\delta}$ for some $d,\tau$
and $\delta$. Using obvious notations, we find that (\ref{eqn:thekey}) is
replaced by
\begin{equs}
\HALF \partial_t\|f\|_{\Ltwo}^2&\leq
 \|A^{1/2}f\|_{\Ltwo}^2
-\|A^{2}f\|_{\Ltwo}^2
+\|Af\|_{\Ltwo}^2
+{\textstyle\frac{\sqrt{\pi}}{2}}
\|(1+A^{2})f\|_{\Ltwo}
\|f\|_{\Ltwo}\|g\|_{\Ltwo}
+c_{per}
\|Af\|_{\Ltwo}\|f\|_{\Ltwo}~,
\end{equs}
from which it follows that
\begin{equs}
\partial_t\|f\|_{\Ltwo}^2 \leq
(c_*+\|g\|_{\Ltwo}^2)
\|f\|_{\Ltwo}^2~,
\end{equs}
for some positive $c_*$. The remainder of the proof is straightforward.

\section{Proof of Theorem \ref{thm:main}}
Let now $\tau_0=0$ and $\nu(x,\tau_0)=\nu_0(x)$ with $\nu_0\in\L^2$,
and define $\sigma_0=\|\nu_0\|_{\Ltwo}$. From the results of the
preceding section, we know that there exists a unique solution of
(\ref{eqn:KSmodifrap}) in ${\cal B}_{d_0}\subset{\cal
W}_{\tau_0,\delta_0 }$ for $d_0=D_*\sigma_0$ and
$\delta_0=C_*\sigma_0^{-2}$, with $C_*$ so small that $\delta_0
<\delta _*$. Let $\tau_1=\tau_0+\delta_0 $. By the definition of
${\cal W}_{\tau,\deltat}$ (see also Corollary \ref{cor:tendtozero}),
$\nu$ and all its derivatives tend to $0$ as $x\to\pm\infty$ for all
$t\in[\tau_0,\tau_1]$. {\em In particular, the trilinear form
$\int\nu(\nu^2)'$ satisfies
$\int\nu(\nu^2)'=-\frac{1}{3}\int(\nu^3)'=0$.} Hence, multiplying
(\ref{eqn:KSmodif}) with $\nu$ and integrating over the space
variable, we get:
\begin{equs}
\HALF \partial_t\int\nu^2&=
-\int\nu\nu''''
-\int\nu\nu''
-\HALF \int\nu(\nu^2)'
-\int\nu(\nu{\psi}_{a})'
-\int\nu\chitzero \\
&=
-\int(\nu'')^2
+\int(\nu')^2
-\HALF \int\nu^2\psitzero '
-\int\nu\chitzero ~.
\end{equs}
Note that because the trilinear form vanishes, we get only quadratic
(and linear) terms in $\nu$, and therefore it is natural to find an
exponential bound in time for the evolution of the $\L^2$ norm; this
is the main explanation for the bounds which follow below. 
Using $-k^4+k^2\le \frac{1}{4}$, we have the inequalities
\begin{equs}
-\int(\nu'')^2
+\int(\nu')^2&\le {\textstyle{\frac{1}{4}}}\int \nu^2~, \\
|\psi_a'| &\le a~, \\
\bigg|\int \nu \chi_a \bigg| &\le \HALF \int \nu^2 +\HALF \int \chi_a^2~.
\end{equs}
We find for,
$t\in[\tau _0,\tau _1]$,  with $\beta \equiv 2(\frac{3}{4}+a)$,
\begin{equs}
\frac{1}{2} \partial_t\int\nu^2&
\,\le\,\frac{\beta }{2}
\int\nu^2+
\frac{1}{2} \int\chitzero ^2
\,\le\,{\frac{\beta }{2}}
\int\nu^2+
\frac{B^2}{2}~.
\label{eqn:lineq}
\end{equs}
This differential inequality is valid for all $t\in[\tau_0,\tau_1]$,
and implies that
\begin{equs}
\|\nu(\cdot,\tau_1)\|_{\Ltwo}
\leq\ed^{\beta \tau _1/2}\sqrt{\sigma_0^2+B^2}\equiv\sigma_1~.
\end{equs}
Again, from the results of the preceding section, we now see that
there exists a unique solution of (\ref{eqn:KSmodifrap}) in ${\cal
B}_{d_1}\subset{\cal W}_{\tau_1,\deltat _1}$ for $d_1=D_*\sigma_1$ and
$\delta_1=C_*\sigma_1^{-2}$. Thus (\ref{eqn:lineq}) is valid for all
$t\in[\tau_0,\tau_2]$ with $\tau_2=\tau_1+\delta_1 $, and we get
\begin{equs}
\|\nu(\cdot,\tau_2)\|_{\Ltwo}
\leq\ed^{\beta \tau_2/2}\sqrt{\sigma_0^2+B^2}\equiv\sigma_2~.
\end{equs}
Continuing by induction, we find
\begin{equs}
\delta _n=C_*\sigma_n^{-2}=\frac{\ed^{-\beta\tau_n}}{\sigma_0^2+B^2}~,
\end{equs}
so that 
\begin{equs}
\tau _{n+1}= \tau _n +E_* \ed^{-\beta\tau_n}~,
\end{equs}
with $E_*=(\sigma_0^2+B^2)^{-1}$. This implies that $\lim_{n\to\infty
} \tau _n =\infty $, and therefore (\ref{eqn:lineq}) is valid for all
$t>0$. This completes the proof of Theorem \ref{thm:main} for the
profile case. The periodic case follows along the same lines.

\subsection*{Acknowledgements}This research was partially supported by the
Fonds National Suisse.

\bibliographystyle{JPE}
\markboth{\sc \refname}{\sc \refname}
\bibliography{refs}

\end{document}